\documentstyle[seceq,epsf,wrapft]{ptptex}

\def\ggs{\buildrel\textstyle > \over {\hbox{\raise0.2ex\hbox{$\sim$}}}}
\def\lls{\buildrel\textstyle < \over {\hbox{\raise0.2ex\hbox{$\sim$}}}}
\def\gsim{\,\lower0.75ex\hbox{$\ggs$}\,}
\def\lsim{\,\lower0.75ex\hbox{$\lls$}\,}
\def \virg{\;\;,}
\def \point{\;\;.}
\def \d{{\rm d}}
\def \e{{\rm e}}
\def \exp{{\rm exp}}
\def \kf{k_{\rm F}}
\def \vf{v_{\rm F}}
\def \ng{\widetilde{g}}
\def \ngt{\widetilde{g_2}}
\def \nq{\widetilde{q}}
\def \no{\widetilde{\omega}}
\def \nv{\widetilde{v}}
\def \nvc{\widetilde{v}_{\rho}}
\def \nvs{\widetilde{v}_{\sigma}}
\def \nT{\widetilde{T}}
\def \nA{\widetilde{A}_+(\widetilde{q},\widetilde{\omega})}
\def \nAs{\widetilde{A}_{+,\sigma}}
\def \nAr{\widetilde{A}_{+,\rho}}
\def \vc{v_{\rho}}
\def \vs{v_{\sigma}}
\def \step{{\rm \theta}}
\def \gac{\gamma_{\rho}}
\def \gas{\gamma_{\sigma}}
\def \virg{\;\;,}
\def \point{\;\,.}
\def \kf{k_{\rm F}}
\def \vf{v_{\rm F}}

\def \e{{\rm e }}

\def \d{{\rm d}}

\def \e{{\rm e}}
\def \exp{{\rm exp}}

\def \intinf{\int_{- \infty}^{\infty}}
\def \kf{k_{\rm F}}
\def \vf{v_{\rm F}}

\notypesetlogo  
\title{
 Effect of Thermal Fluctuation on 
  Spectral Function for 
 the Tomonaga-Luttinger Model\footnote{Submitted to Prog. Theor. Phys. } 
}

\author{
 Naoki {\sc Nakamura} and Yoshikazu {\sc Suzumura}\\ 
}

\inst{
Department of Physics, Nagoya University,
Nagoya 464-01 
}

\recdate{
 May, 10 1997
}

\abst{%
 We examine 
 the spectral function of the single electron Green function 
 at finite temperatures 
  for the  Tomonaga-Luttinger  model 
 which consists of  the mutual interaction with  only  
  the forward scattering.  
  The   spectral weight, which is calculated 
  as a function of the frequency 
  with  the fixed  wave number, shows that  
   several peaks originating in the excitation spectra of 
    charge and spin   fluctuations    
  vary  into a single peak by the increase of temperature. 
}
\begin{document}
\maketitle
\section{Introduction}

 The Tomonaga-Luttinger model\cite{Tomonaga,Luttinger}, 
 which can be solved exactly,  is  a fundamental model 
  for  studying  the electronic state 
  in one-dimensional interacting electron systems. 
 Since the  excitations are  gapless  for  both charge and spin 
  degree of   freedom, 
 the response functions for 
  the  superconducting state, spin density wave state and charge 
   density wave state show the power law behavior with respect 
   to  frequency and temperature 
 where the exponents depend on the interactions.
\cite{Luther,Solyom} 
 
 Noticeable   properties  of the Tomonaga-Luttinger model 
 have been obtained 
  in  the state of the single electron. 
 The power law dependence  has been found  
  in the momentum distribution function
  \cite{Mattis} 
   around the Fermi momentum and 
  the density of states
\cite{Suzumura_P2,Nakamura,Voit} 
   around the Fermi energy. 
 These facts are characteristic of 
   the Luttinger liquid
 \cite{Haldane}   where 
    the quantum fluctuation in one-dimension leads to 
  the  marginal behavior 
  between the metallic  state and the ordered state.   
  The detail  of the electronic  state 
 is obtained  in the  spectral function  
  which is calculated from  
   the imaginary part of the single electron Green function. 
  The spectral function 
 at absolute zero temperature exhibits    
    several  peaks  
 which are associated with 
 the fluctuation of the  pairing  electron 
 and the separation of the charge and spin degrees of freedom 
in the presence of the interaction.
\cite{Meden,Voit} 
   The spectral function    for the Hubbard model 
  with the quarter-filled band 
   and the infinite   repulsive interaction
\cite{Penc2} 
 has been calculated 
 where  the frequency-dependence is similar to 
 that of  the Luttinger liquid. 
 However  
   the temperature-dependence  of   the spectral function 
 of the Tomonaga-Luttinger model   is not studied qualitatively 
   although     the thermal fluctuation 
    takes an important role for the electronic state 
     around the Fermi energy.
\cite{Suzumura_P2,Nakamura}    

  In the present paper, we examine 
  the spectral function at finite temperatures 
   to understand   how  several peaks 
    are varied by the thermal fluctuation. 
 In \S2, formulation is given in terms of the single electron 
 Green function at finite temperatures. 
 In \S3, the frequency-dependence of 
 the spectral function is examined numerically 
 with some choices of interaction and momentum. 
 The  crossover from the ground state 
  to the  state expected at finite temperature is demonstrated. 
 \S 4 is devoted to  discussion. 
\section{Formulation}
%
 The Tomonaga-Luttinger model is given by 
\begin{eqnarray}
                                                 \label{eqn:FH}
 H &=& \sum_{r = \pm}
       \sum_{ s = \uparrow, \downarrow}
       \sum_{k}
         \vf ( r k - \kf )
         C_{k , s , r}^+ C_{k , s , r}
 \nonumber\\
    && + \frac{\pi \vf}{2L}
         \sum_{r,s,s'}
         \sum_{k_1 , k_2 , q}
          \e^{- |q| \Lambda}
          \biggl[
           \left(
             \widetilde {g}_{2 \parallel} \delta_{s , s'}
            +\widetilde {g}_{2 \perp} \delta_{s , -s'}
           \right)
           C_{k_1 , s , r}^+  C_{k_2 , s' , -r}^+
           C_{k_2+q , s' , -r} C_{k_1-q , s , r} 
                                 \nonumber  \\
       && + \widetilde {g}_{4 \perp} \delta_{s , -s'}
           \left[
            C_{k_1 , s , r}^+ C_{k_2 , s' , r}^+
            C_{k_2+q , s' , r} C_{k_1-q , s , r}
           \right]  \biggl]
  \virg
\end{eqnarray}
 where $C_{k,s,r}^{+}$ denotes the creation operator of 
the fermion  with spin $s=\uparrow(\downarrow) (=+(-))$ 
and the momentum $k$ 
being positive (negative) for $r=+(-)$.   
 The first term is the kinetic energy where 
$\vf$ and $\kf$ are Fermi velocity  and Fermi momentum respectively. 
 In  the second term, 
the quantity 
   $\widetilde {g}_{2 \parallel, \perp}$   
   $( \equiv g_{2 \parallel, \perp}/ \pi \vf )$ 
denotes the normalized coupling constant of the interaction 
for the forward scattering between 
two kinds of electrons with $r = +$ and  $ r= -$ 
   and  $\widetilde {g}_{4 \perp }$   $( \equiv g_{4 \perp}/\pi \vf )$ 
 is that for  the same kind of electrons.  
 Quantities  $\Lambda^{-1}$ and $L$ denote      
 the  momentum cutoff of the interaction and  
 the length of the system respectively.  
 
 Based on the bosonization method, \cite{Mattis} 
eq. (\ref{eqn:FH}) 
 in case of $|q| \lsim \Lambda^{-1}$ 
 is expressed as \cite{Suzumura_P1} 
\begin{equation}
                                                   \label{eqn:PH}
 H_P = \sum_{\nu = \rho , \sigma}
        \frac{v_{\nu}}{4 \pi}
        \int_{- \infty}^{\infty} \d x
         \left[
          \frac1{ \eta_{\nu}}
           \left(
            \partial_x \theta_{\nu , +} (x)
           \right)^2
          +  \eta_{\nu}
          \left(
           \theta_{\nu , -} (x)
          \right)^2
         \right]
 \virg
\end{equation}
\begin{eqnarray}
                                                    \label{eqn:v}
 v_{\nu}
  &=& \vf \sqrt{
            \left(
             1 \pm  \widetilde {g}_{4 \perp}/2
            \right)^2
          - \left(
            (    \widetilde {g}_{2 \parallel}
             \pm \widetilde {g}_{2 \perp})/2
           \right)^2
          }
 \virg\\
                                                  \label{eqn:eta}
 \eta_{\nu}
  &=& \sqrt{
       \frac{
        1 \pm \widetilde {g}_{4 \perp}/2
        - (    \widetilde {g}_{2 \parallel}
           \pm \widetilde {g}_{2 \perp}
          )/2
       }
       {
        1 \pm \widetilde {g}_{4 \perp}/2
        + (    \widetilde {g}_{2 \parallel}
           \pm \widetilde {g}_{2 \perp}
          )/2
       }
      }
 \virg
\end{eqnarray}
where $+$($-$) corresponds to  $\rho$($\sigma$) 
 and
\begin{eqnarray}
                                             \label{eqn:trho}
 \theta_{\rho , \pm} (x)
 &=& \frac{i}{2}
     \sum_{q}
      \frac{2 \pi}{L p}
      \e^{- \frac{\alpha_0}2 | q | - i q x}
     \sum_{k , s}
      \left[
       C_{k+q , s , +}^+ C_{k , s , +}
       \pm 
       C_{k+q , s , -}^+ C_{k , s , -}
      \right]
 ,\\
                                           \label{eqn:tsigma} 
 \theta_{\sigma , \pm} (x)
 &=& \frac{i}{2}
     \sum_{q}
      \frac{2 \pi}{L q}
      \e^{- \frac{\alpha_0}2 | q | - i q x}
     \sum_{k , s}s
      \left[
       C_{k+q , s , +}^+ C_{k , s , +}
       \pm 
       C_{k+q , s , -}^+ C_{k , s , -}
      \right]
 .
\end{eqnarray}
Equations (\ref{eqn:trho}) and (\ref{eqn:tsigma}), 
 which denote the phase variables for the charge and 
  spin fluctuations respectively, 
  satisfy the commutation relation given by  
\begin{eqnarray}
                                                  \label{eqn:com}
 \left[
  \theta_{\nu ,  \pm} (x) ,
    (-1 /2 \pi)\partial_{x'}\theta_{\nu' , \mp} (x')
 \right]
  = i \delta_{\nu , \nu'}
      \delta (x - x') \point 
\end{eqnarray}
  From  eqs. (\ref{eqn:PH}) and (\ref{eqn:com}),
   one obtains the excitation spectra,
    $\vc q$ and $\vs q$ for the charge fluctuation 
     and the spin fluctuation respectively. 
 By introducing the cutoff parameter 
    $\alpha_0 (\rightarrow +0)$  for the convergence and  
  making use of the fermion field operator 
defined as\cite{Luther}
\begin{equation}
                                                \label{eqn:Field}
 \mit\Psi_{s , r}^P (x ,t)
  = \frac{ \e^{r i \kf x} }{ \sqrt{2 \pi \alpha_0} }
    \exp \left[
          \frac{r i}2
              \left[
                 \left(
                      \theta_{\rho ,+} (x,t)
                  + r \theta_{\rho ,-} (x,t)
                 \right)
               + s
                 \left(
                      \theta_{\sigma ,+} (x,t)
                  + r \theta_{\sigma ,-} (x,t)
                 \right)
              \right]
         \right]
 ,
\end{equation}
 the retarded Green function at finite temperatures is 
  calculated as\cite{Suzumura_P2}
\begin{eqnarray}
                                                \label{eqn:Green}
 G_{s , r}^R (x , t)
  &\equiv& -i \theta (t)
      \left<
          \mit\Psi_{s,r}^P (x,t)     \mit\Psi_{s,r}^P (0,0)^{+}
        + \mit\Psi_{s,r}^P (0,0)^{+} \mit\Psi_{s,r}^P (x,t)
      \right>_{H_P}    \nonumber \\
  &=& -i \frac{ \theta(t) }{2 \pi} e^{i r \kf x}
         \left[
          \prod_{\nu = \rho , \sigma}
           \left[
            \frac1
            { ( \alpha_0 + i (v_{\nu} t - r x) )^{\frac12} }
            \left(
             \frac{ \Lambda^2 }
             { \left( \Lambda + i v_{\nu} t \right)^2 + x^2 }
            \right)^{\gamma_{\nu}}
           \right]
         \right.
 \nonumber \\
   &&    \left.
          \times
          \Xi (x , t , T)
          + {{x \to -x} \choose {t \to -t}}
         \right]
 \virg
\end{eqnarray}
\begin{eqnarray}
                                                   \label{eqn:Xi}
 &&\Xi (x , t , T)
 \nonumber\\
 && = \prod_{\nu = \rho , \sigma}
    \prod_{n = 1}^{\infty}
     \Biggl[
        \biggl[
         1 +
         \Bigl(
          \frac{ v_{\nu} t - r x }
          { \alpha_0 + \frac{n v_{\nu}}{T} }
         \Bigr)^2
        \biggr]^{\frac12}
        \biggl[
         1 +
         \Bigl(
          \frac{ v_{\nu} t - r x }
          { \Lambda + \frac{n v_{\nu}}{T} }
         \Bigr)^2
        \biggr]^{\gamma_{\nu}}
        \biggl[
         1 +
         \Bigl(
          \frac{ v_{\nu} t + r x }
          { \Lambda + \frac{n v_{\nu}}{T} }
         \Bigr)^2
        \biggr]^{\gamma_{\nu}}
     \Biggr]^{-1}.
 \nonumber                                                       \\
\end{eqnarray}
 The quantity $\gamma_{\nu}$,  which denotes the 
  magnitude of the interaction, is defined by  
\begin{equation}
                     \label{eqn:gamma}
 \gamma_{\nu} = (\eta_{\nu} + \eta_{\nu}^{-1} - 2) /8
 \point
\end{equation}
 The quantity, $T$, is the temperature and  
$k_{\rm B}$ is taken as unity. 
In deriving eq. (\ref{eqn:Green}) from eq. (\ref{eqn:Field}), 
we have used an approximation 
that the second line of eq. (\ref{eqn:Green}) 
is correct in case of 
$|x|, |\vf t| \gg \Lambda$ 
and results in the extra  factor of
$\vf / (\vc \vs)^{1/2}$ 
 in case of 
$|x|, |\vf t| \ll \Lambda$ 
compared with the exact one.\cite{Dzyaloshinskii,Suzumura_P2} 
Actually, by noting that  $\Xi (x , t , T) \rightarrow 1$ 
 in the limit of absolute zero temperature, 
eq. (\ref{eqn:Green}) in case of $T$ = 0
becomes equal to the retarded Green function 
 obtained by Luther and Peschel 
 \cite{Luther,Meden,Voit}
 which is valid for the length scale 
  being larger than $\Lambda$. 
Therefore 
the following calculation of the spectral function is justified  
when  the frequency ( the momentum )
is smaller than $\vf / \Lambda$ ($ \Lambda^{-1}$). 
 We note the difference between  
 two kinds of cutoff parameters 
  in the Tomonaga-Luttinger model.  
   The quantity $\Lambda$ leading to  
     the   momentum  cutoff  of the interaction 
 plays  an  essential role for  the existence of 
  the  characteristic energy, $\vf/\Lambda$,  
 but the quantity $\alpha_0$  vanishes 
 in the end  by  taking the limit of 
$\alpha_0 \to 0$. 

%
    
 In terms of eq. (\ref{eqn:Green}), 
the spectral function is calculated as 
\begin{eqnarray}
                                           \label{eqn:spectral}
 A_r (q , \omega)
  &=&  - \frac1{\pi} {\rm Im}
       \left[
        \int_{- \infty}^{\infty} \d x
        \int_{- \infty}^{\infty} \d t
         \e^{- i \{(q+r\kf) x - \omega t\}}
         G_r^R (x , t)
       \right]
 \nonumber \\
 &=& \frac1{(2 \pi)^2}
      \int_{- \infty}^{\infty} \d x \int_{- \infty}^{\infty} \d t
       \biggl[
        \e^{-i (q x - \omega t)}
 \nonumber\\
 &&    \times
        \prod_{\nu = \rho , \sigma}
         \biggl[
          \frac1{ (\alpha_0 + i (v_{\nu} t - r x))^{\frac12} }
          \Bigl(
           \frac{ \Lambda^2 }
           {
            \left(
             \Lambda + i v_{\nu} t
            \right)^2
            +
            x^2
           }
          \Bigr)^{\gamma_{\nu}}
         \biggr]
       \Xi (x , t , T)
 \nonumber\\
 &&    + {{\omega \to - \omega} \choose {q \to -q}}
       \biggr]
 \point
\end{eqnarray}

We examine eq. (\ref{eqn:spectral}) for  two kinds of 
 cases.   
 The case (i) is given by 
 $\widetilde{g}_{2\parallel} \equiv \ng \not=0$ and  
 $\widetilde{g}_{2 \perp} 
 =\widetilde{g}_{4 \perp}=0$ which corresponds to 
 the spinless  Tomonaga-Luttinger model. 
  The case (ii) is given by 
 $\widetilde{g}_{2 \parallel} = \widetilde{g}_{2 \perp} 
 = \widetilde{g}_{4 \perp} \equiv 
  \widetilde{g} \not = 0$ 
 which represents  the spinful Tomonaga-Luttinger model. 
%
%

 In the case (i),  eq. (\ref{eqn:spectral}) is rewritten   as 
\begin{equation}
                                            \label{eqn:spectral1}
 A_r( q , \omega) = \frac1{8 {\pi}^2 v}
                     \left[
                      F_1 \left( \Omega_1 , T \right)
                      F_2 \left( \Omega_2 , T \right)
                     + {{\Omega_1 \to -\Omega_1}
                         \choose
                        {\Omega_2 \to -\Omega_2}}
                     \right]
 \point
\end{equation}
 By use of 
$
 \Omega_1 = \left(
             \omega + r {v} q
            \right)/2 v
$, 
$
 \Omega_2 = \left(
             \omega - r v q
            \right)/2 v
$,
$ s_1 = v t - r x$ and
$ s_2 = v t + r x$,   quantities 
 $F_1(\Omega_1,T)$ and $F_2(\Omega_2,T)$ are written as 
\begin{eqnarray}
                                                   \label{eqn:F1}
 &&F_1 ( \Omega_1 , T )
 \nonumber\\
 &&= \int_{- \infty}^{\infty} \d s_1
       \frac{ \e^{i \Omega_1 s_1} }{ \alpha_0 + i s_1 }
       \left(
         \frac{\Lambda}{\Lambda + i s_1}
       \right)^{\gamma}
     \times
       \prod_{n = 1}^{\infty}
        \Biggl[
         \biggl[
          1 +
          \Bigl(
           \frac{ s_1  }{ \alpha_0 + \frac{n v}{T} }
          \Bigr)^2
         \biggr]
         \biggl[
          1 + \Bigl(
               \frac{s_1}{\Lambda + \frac{n v}T}
              \Bigr)^2
         \biggr]^{\gamma}
        \Biggr]^{-1}
 \nonumber \\
 &&= \pi +
        2 \Lambda^{\gamma}
        \int_0^{\infty} \d s_1
         \frac{
          \sin \left[
                  \Omega_1 s_1
                - \gamma 
                  \tan^{-1} \left( 
                             \frac{s_1}{\Lambda}
                            \right)
               \right]
         }
         {s_1
          \left( \Lambda^2 + s_1^2 \right)^{ \frac{\gamma}2 }
         }
         \frac{
          T s_1 / v
         }
         {
          \sinh \left(
                 T s_1 / v
                \right)
         }
          \prod_{n = 1}^{\infty}
           \biggl[
            1 + \Bigl(
                 \frac{s_1}{\Lambda + \frac{n v}T}
                \Bigr)^2
           \biggr]^{- \gamma},
 \nonumber\\ \\
                                                   \label{eqn:F2}
 &&F_2 ( \Omega_2 , T )
 \nonumber\\
  &&= \int_{- \infty}^{\infty} \d s_2
       \e^{i \Omega_2 s_2}
       \left(
         \frac{\Lambda}{\Lambda + i s_2}
       \right)^{\gamma}
      \prod_{n = 1}^{\infty}
        \biggl[
         1 + \Bigl(
              \frac{s_2}{\Lambda + \frac{n v}T}
             \Bigr)^2
        \biggr]^{- \gamma}
     \nonumber \\
 &&= 2 \Lambda^{\gamma}
       \int_0^{\infty} \d s_2
        \frac
        {
         \cos \left[
                 \Omega_2 s_2 
               - \gamma \tan^{-1} \left(
                                   \frac{s_2}{\Lambda}
                                  \right)
              \right]
        }
        {
         \left( \Lambda^2 + s_2^2 \right)^{\frac{\gamma}2}
        }
        \prod_{n = 1}^{\infty}
         \biggl[
          1 + \Bigl(
               \frac{s_2}{\Lambda + n \frac{v}T}
              \Bigr)^2
         \biggr]^{- \gamma},
\end{eqnarray}
 where  
   $v=v_{\rho}=v_{\sigma} = \vf [1-(\widetilde{g}_{2}/2)^2]^{1/2}$ and 
 $\gamma = [ (1 - (\ngt/2)^2)^{-1/2} -1]/2$. 

 In the case (ii), eq. (\ref{eqn:spectral}) is rewritten as 
\begin{eqnarray}
                                            \label{eqn:spectral2}
 A_r (q , \omega)
 &=&
  \frac1
  { (2 \pi)^2 |v_{\rho} - v_{\sigma}| }
   \intinf \d s_{\rho}
   \intinf \d s_{\sigma}
    \Bigl[
     \e^{i(\Omega_{\sigma} s_{\rho} + \Omega_{\rho} s_{\sigma})}
 \nonumber\\
 &&    \times
        \prod_{\nu = \rho , \sigma}
         \left[
          \frac1{ (\alpha_0 + i s_{\nu})^{\frac12} }
          \left(
           \frac{ \Lambda }
           { \Lambda + i s_{\nu} }
          \right)^{\gamma_{\nu}}
          \left(
           \frac{ \Lambda }
           { \Lambda + i (a_{\nu} s_{\rho}+b_{\nu} s_{\sigma}) }
          \right)^{\gamma_{\nu}}
         \right]
       \Xi (x , t , T)
 \nonumber\\
    &&+
      { { \Omega_{\rho}   \to - \Omega_{\rho} }
                       \choose
        { \Omega_{\sigma} \to - \Omega_{\sigma} } }
    \Bigr]
 \nonumber\\
 &=&
  \frac{ \Lambda^{2 (\gamma_{\rho} + \gamma_{\sigma}) } }
  { (2 \pi)^2 |v_{\rho} - v_{\sigma}| }
   \left[
    \left[
       F_s (\Omega_{\rho},\Omega_{\sigma},T)
     + F_c (\Omega_{\rho},\Omega_{\sigma},T)
    \right]
     +
      { { \Omega_{\rho}   \to - \Omega_{\rho} }
        \choose
        { \Omega_{\sigma} \to - \Omega_{\sigma} } }
   \right]
 \point
 \nonumber\\
\end{eqnarray}
 By use of 
$
 \Omega_{\sigma}
  = (   \omega - r v_{\sigma} q)/(v_{\rho} -v_{\sigma}),
 \ 
 \Omega_{\rho}
  = (r v_{\rho} q - \omega)/(v_{\rho} - v_{\sigma}),
 \ 
 s_{\rho} = v_{\rho} t - r x,
 \ 
 s_{\sigma} = v_{\sigma} t - r x,
 \ 
 a_{\rho}
  = (v_{\rho} + v_{\sigma})/(v_{\rho} - v_{\sigma}),
 \ 
 b_{\rho}
   =  - 2 v_{\rho}/(v_{\rho} - v_{\sigma}),
 \ 
  a_{\sigma}
  = 2 v_{\sigma} / ( v_{\rho} - v_{\sigma} ),
 \ 
 b_{\sigma}  =  - (v_{\rho} + v_{\sigma})/(v_{\rho} - v_{\sigma})
 \ 
 $, 
  quantities 
 $ F_s(\Omega_{\rho},\Omega_{\sigma},T)$ and 
 $ F_c(\Omega_{\rho},\Omega_{\sigma},T)$ are expressed as  
 
\begin{eqnarray}
                                                   \label{eqn:Fs}
 F_s(\Omega_{\rho},\Omega_{\sigma},T)
  &=& 2 \int_0^{\infty} \d s_{\rho}
        \int_0^{\infty} \d s_{\sigma}
         \Big[
          \prod_{\nu = \rho , \sigma}
           s_{\nu} (\Lambda^2 + s_{\nu}^2)^{\gamma_{\nu}}
           (  \Lambda^2
            + (  a_{\nu} s_{\rho}
               + b_{\nu} s_{\sigma})^2 )^{\gamma_{\nu}}
         \Big]^{- \frac12}
 \nonumber\\
 &\times&
         \sin
          \Biggl[
            \Omega_{\sigma} s_{\rho} + \Omega_{\rho} s_{\sigma}
           - \sum_{\nu = \rho , \sigma}
              \gamma_{\nu}
              \biggl[
               \tan^{-1}
                \Bigl(
                 \frac{s_{\nu}}{\Lambda}
                \Bigr)
              +\tan^{-1}
               \Bigl(
                \frac{a_{\nu} s_{\rho} + b_{\nu} s_{\sigma}}
                {\Lambda}
               \Bigr)
             \biggr]
          \Biggr]
 \nonumber\\
 &\times&
      \xi (s_{\rho},s_{\sigma},T)
 \virg
\end{eqnarray}
\begin{eqnarray}
                                                   \label{eqn:Fc}
 F_c(\Omega_{\rho},\Omega_{\sigma},T)
 &= & 2 \int_0^{\infty} \d s_{\rho}
       \int_{-\infty}^0 \d s_{\sigma}
        \Bigl[
         \prod_{\nu = \rho , \sigma}
          |s_{\nu}| (\Lambda^2 + s_{\nu}^2)^{\gamma_{\nu}}
          (  \Lambda^2
           + (  a_{\nu} s_{\rho}
              + b_{\nu} s_{\sigma})^2)^{\gamma_{\nu}}
        \Bigr]^{- \frac12}
 \nonumber\\
 &\times&
         \cos
          \Biggr[
           \Omega_{\sigma} s_{\rho} + \Omega_{\rho} s_{\sigma}
           -\sum_{\nu = \rho , \sigma}
             \gamma_{\nu}
             \biggl[
              \tan^{-1}
                \Bigl(
                 \frac{s_{\nu}}{\Lambda}
                \Bigr)
              +\tan^{-1}
               \Bigl(
                \frac{a_{\nu} s_{\rho} + b_{\nu} s_{\sigma}}
                {\Lambda}
               \Bigr)
             \biggr]
          \Biggr]
 \nonumber\\
 &\times&
      \xi (s_{\rho},s_{\sigma},T)
 \virg
\end{eqnarray}
where 
 $\gamma_{\rho} = 
  [ (1 - (  2 \ng/(2+\ng))^2)^{-1/2} -1]/4$, 
 $\gamma_{\sigma} = 0$,
   $ v_{\rho}= \vf [(1+\ng/2)^2 -\ng^2]^{1/2} $ and 
   $ v_{\sigma}=   \vf (1-\ng/2) $.  
 In deriving eqs. (\ref{eqn:Fs}) and (\ref{eqn:Fc}), 
 we made use of the following approximation,
 \cite{Ogata,Nakamura} 
\begin{eqnarray}
                                                \label{eqn:Xiapp}
 \Xi (x,t,T)
  &\simeq&
      \prod_{\nu = \rho , \sigma}
      \left[
       \left[
        \frac{ T s_{\mu} / v_{\nu} }
        { \sinh \left( T s_{\nu} / v_{\nu} \right) }
       \right]^{\frac12 + \gamma_{\nu}}
       \left[
        \frac{T(a_{\nu} s_{\rho} + b_{\nu} s_{\sigma}) / v_{\nu}}
        { \sinh
         \left(
          T (a_{\nu} s_{\rho} + b_{\nu} s_{\sigma}) / v_{\nu}
         \right)
        }
       \right]^{\gamma_{\nu}}
      \right]
           \nonumber \\
 & \equiv &  \xi (s_{\rho} , s_{\sigma} , T)
 \virg 
\end{eqnarray}
%
 which was obtained  
 by discarding  $\Lambda$ in eq. (\ref{eqn:Xi}). 
Such a treatment is valid for 
$T \lsim \vf / \Lambda$ 
in the present calculation. 

  In addition to the case (ii), 
we  examine the case of  $\widetilde{g}_{4 \perp} \not= 0$ 
 and zero otherwise which corresponds to 
 the one-branch Luttinger liquid,
\cite{Voit} 
 i.e., 
 the forward scattering  within the same kind of electrons. 
In this case, 
 the spectral function $A_{+}(q,\omega)$ is obtained 
 by putting $\gamma_{\rho} = \gamma_{\sigma}=0$ in 
eq. (\ref{eqn:spectral2}) 
 where $A_{+}(q,\omega)$  also shows  the separation 
  of the charge and spin degrees of freedom.  
%
%
\section{Spectral Function}
 We evaluate the spectral weight 
  of eq.(\ref{eqn:spectral}) which is 
 normalized as, 
\begin{eqnarray}
                                                \label{eqn:scale}
 \widetilde {A}_{+} (\widetilde {q},\widetilde {\omega})
  = A_{+} (q, \omega) \vf \Lambda^{-1}
 \point  
\end{eqnarray}
 Quantities  $q$, $\omega$, $v$ and $T$ are also normalized as  
$ \widetilde {q} = q \Lambda $, 
$ \widetilde {\omega} = \omega / (\vf \Lambda^{-1}) $,
$ \widetilde{v}=v/\vf $  and 
$ \widetilde {T} = T / (\vf \Lambda^{-1})$ 
respectively. 
 From eq.(\ref{eqn:gamma}),  
   the parameter for the interaction 
    is defined as 
\begin{eqnarray}
                                                \label{eqn:alpha}
 \alpha
  = 2 \left( \gamma_{\rho} + \gamma_{\sigma} \right) 
 \point
\end{eqnarray}
 We examine $\nA$ in detail  by choosing $\alpha$= 0.125 
 which  corresponds to 
the limit of the large repulsive interaction for 
one-dimensional Hubbard model.\cite{Schulz}
  Since  $ A_{r}(\nq, \no) = A_{r}(-\nq,-\no)=A_{-r}(\nq,-\no) $,  we investigate  numerically $\nA$ in case of  $\nq > 0 $.

%
%
\subsection{spinless case}  
 The model in the case of 
  $g_{2 \parallel}\not= 0$ and 
  $g_{2 \perp} = g_{4 \perp} =0 $ is equivalent  to that of 
  the spinless fermion  since   
there is no distinction between the charge fluctuation and 
the spin fluctuation, i.e., 
  $ \gamma_{\rho} = \gamma_{\sigma}$. 
  
 First, $\nA$ with  $\nq=0$ is examined. 

%
\begin{figure}[htb]
 \parbox{\halftext}{
\vspace{-1.5cm}
\epsfxsize = 8.5cm
\vspace{-1.0cm}
\hspace{0.3cm}
\centerline{\epsfbox{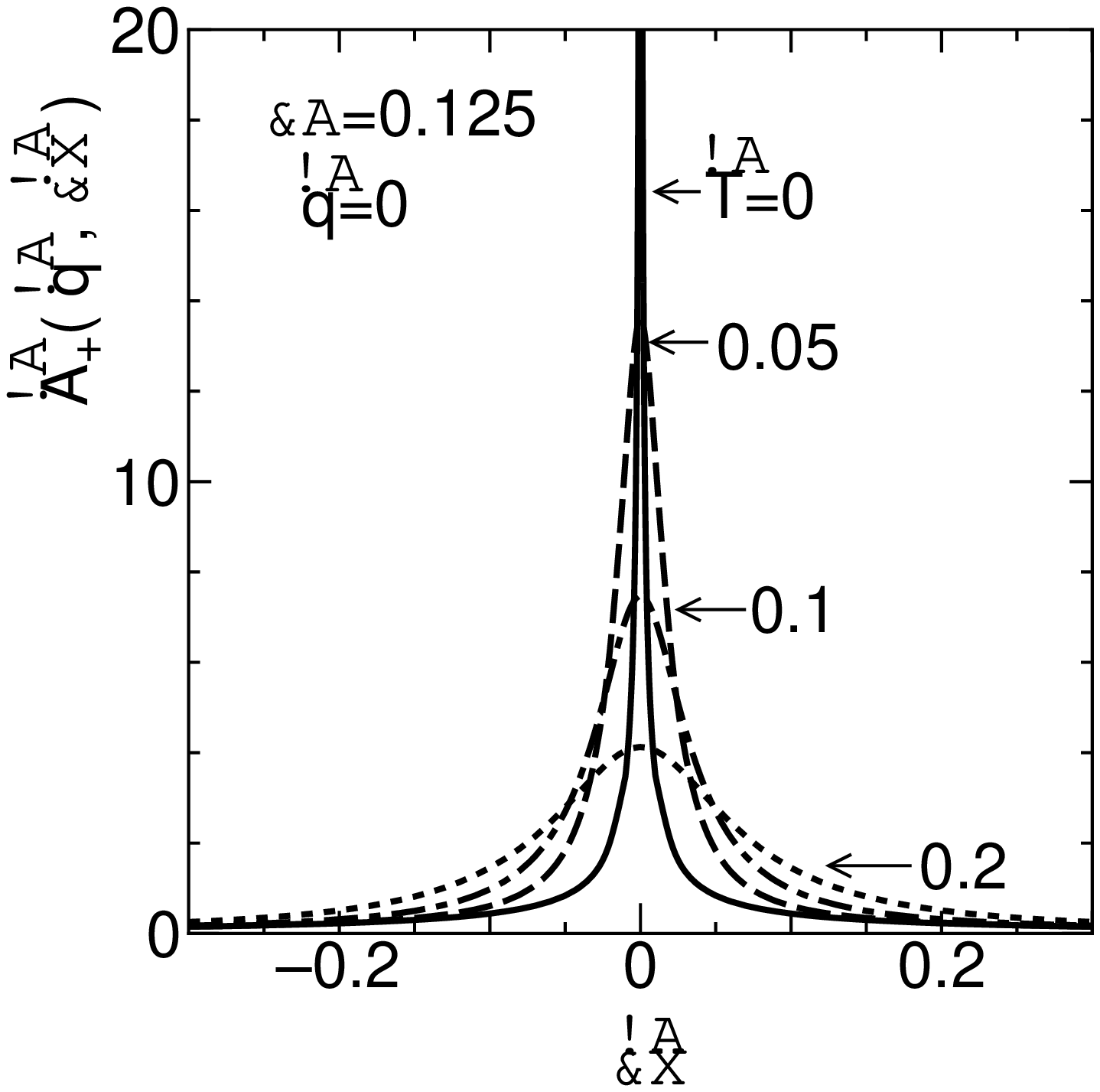}}
\vspace{-2.0cm}
\caption{
 The $\no$-dependence of 
 spectral function  $\nA$ 
  of the spinless model   
   where  $\nq=0$,  $\alpha=0.125$ and $\nT$ is chosen as   
 $\nT$=0, 0.05, 0.1 and 0.2. 
}
\label{fig:1}
}
\hspace{7mm}
\parbox{\halftext}{
\epsfxsize = 8.5cm
\vspace{-1.25cm}
\centerline{\epsfbox{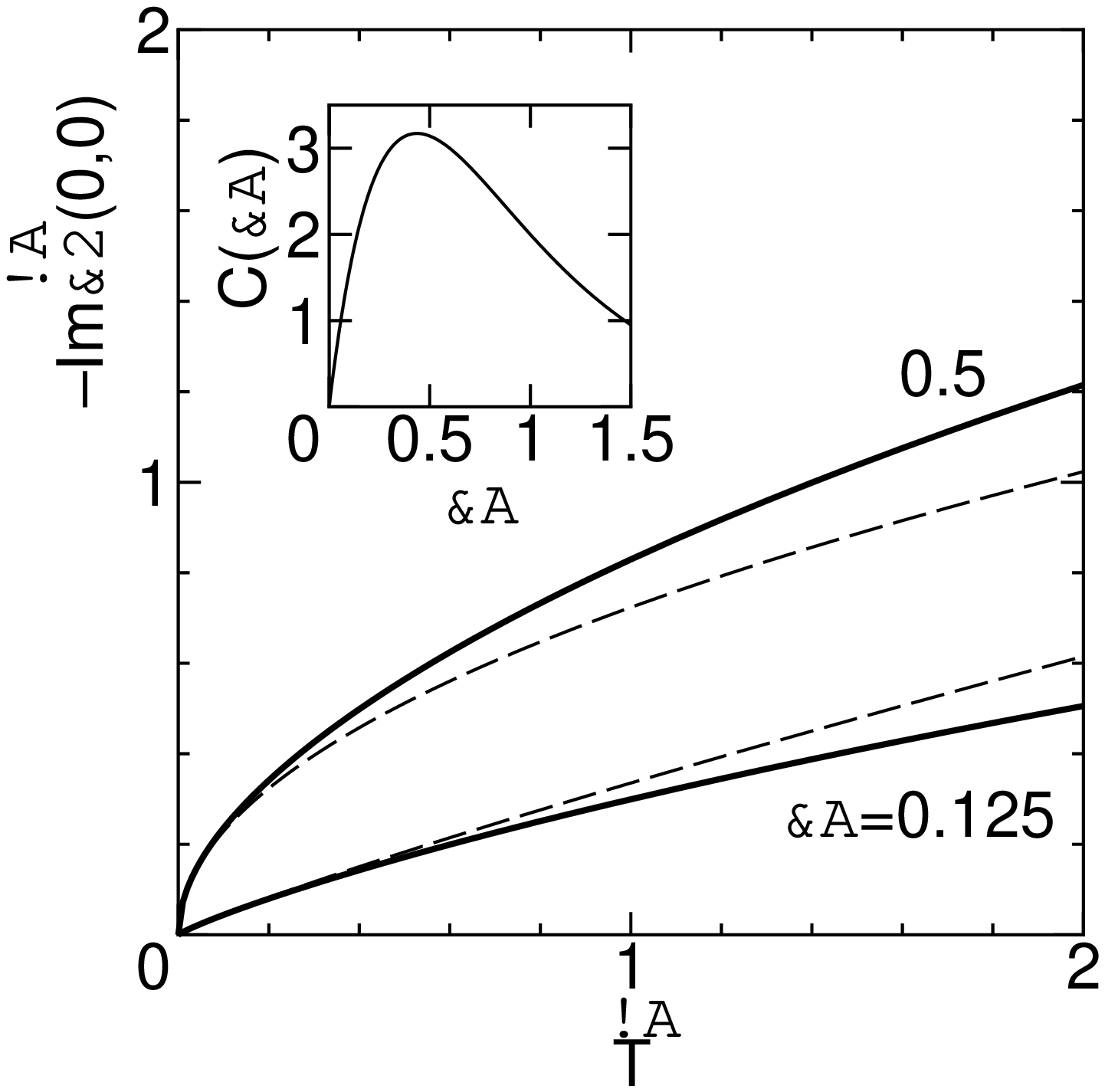}}
\vspace{-1.5cm}
\caption{
 The $\nT$-dependence of 
  the  imaginary part of the normalized self-energy, 
  $ - {\rm Im} \widetilde{\Sigma}^{R} (0,0)$
  $ (=  - {\rm Im} \Sigma^{R} (0,0) \Lambda /\vf)$
   is shown in the case of 
  $\alpha$ = 0.125 and 1.5.   
  The dashed curve is the asymptotic value given by 
   eq. (\protect \ref{eqn:IMSE}) 
  and the  inset shows the coefficient given by 
  eq. (\protect \ref{eqn:Ca}).  
}
\label{fig:2}
}
\end{figure}
%
 In Fig. 1, 
 $\nA$ with  $\alpha$ = 0.125 and $\nq =0$ 
is shown as a function of $\no$  with some choices of $\nT$. 
  In case of $T = 0$,  the spectral function $\nA$  
  diverges at $\no=0$ where  
 $\nA$ with  $\nq=0$ and small $|\no|$ is given by 
  $ \nA \propto \no^{\alpha -1}$.\cite{Meden,Voit}
 For $\nT \not= 0$,  $\nA$ with  $\no = 0$ becomes finite 
  and the peak of $\nA$  decreases and the width increases 
  due to the thermal fluctuation. 
 The half width of $\nA$, which is defined 
  by  $\Delta \no$,     is proportional to   $\nT$ 
    within the numerical accuracy of the present calculation, 
  e.g.  $\Delta \no \sim 0.4 \nT$ ($\sim 2 \nT$ )
    for $\nT < 0.1$  in case of   $\alpha = 0.125 (=0.5)$.  
 The height of the peak at low temperatures is given by 
   $\widetilde{A}_+(0,0) \propto \nT^{\alpha -1}$ 
  which is related to the imaginary part 
  of  the self-energy of the   Fourier transform 
   of the retarded Green function,  eq. (\ref{eqn:Green}). 
  Actually, by defining ${\rm Im} \Sigma^{R}(q,\omega)$ as 
   the imaginary part of the Green function, 
 one finds the relation that 
  $ {\rm Im} \Sigma^{R}(0,0) = - [ \pi A(0,0)]^{-1}$ 
  where 
  $ {\rm Re} \Sigma^{R}(0,0) = 0$.  
 From eq. (\ref{eqn:spectral1}),  
  the quantity $ {\rm Im} \Sigma^{R}(0,0) $ 
 in case of $T \ll v \Lambda^{-1}$ 
  is calculated as
%
\begin{eqnarray}
                                                 \label{eqn:IMSE}
 - {\rm Im} \Sigma^R (0,0)
  & \simeq C(\alpha) &
           \left(
            \frac{T \Lambda}{v}
           \right)^{1 - \alpha}
                         \virg\\
                                                   \label{eqn:Ca}
 C(\alpha)^{-1}
  &=& \frac{
       \pi^{\gamma - 1}
       \sin \left( \pi \gamma \right)
      }{2 v}
      \int_0^{\infty} \d y
       \left[
        \left(
         \frac{\pi}{\sinh(\pi y)}
        \right)^{\gamma+1}
        - y^{-(\gamma + 1)}
       \right]
 \nonumber\\
  &&  \times
      \int_0^{\infty} \d y
       \left(
        \sinh(\pi y)
       \right)^{-\gamma}
 \point
\end{eqnarray}
%
 In Fig. 2,
 the normalized quantity of   ${\rm Im} \Sigma^{R} (0,0)$ 
as a function of $\nT$  is shown by the solid curves 
 for $\alpha = $ 0.125 and 0.5. 
 The dashed curve denotes the asymptotic value given 
  by eq. (\ref{eqn:IMSE}) 
where the good coincidence between the solid curve and the 
dashed curve is  obtained at low temperatures. 
  The  actual value of  ${\rm Im} \Sigma^{R} (0,0)$
  is smaller (larger ) than the  asymptotic one value 
   where  the crossover takes place 
    around $\alpha \simeq  0.32$   at low temperatures. 
 In the inset, $C(\alpha)$ of eq. (\ref{eqn:Ca}) 
is shown as a function of $\alpha$.
 The quantity  $C(\alpha)$ takes a maximum around 
  $\alpha \simeq  0.44$ and   $C(\alpha)$ becomes zero 
   at $\alpha =  0$, 
 indicating the fact  
 that  $A_{r}(q,\omega) \to \delta(\omega-\vf q)$
   at $\alpha \rightarrow 0$. 

%
\begin{wrapfigure}{r}{6.6cm}
\epsfxsize= 8.5 cm
\vspace{-1.5cm}
\centerline{\epsfbox{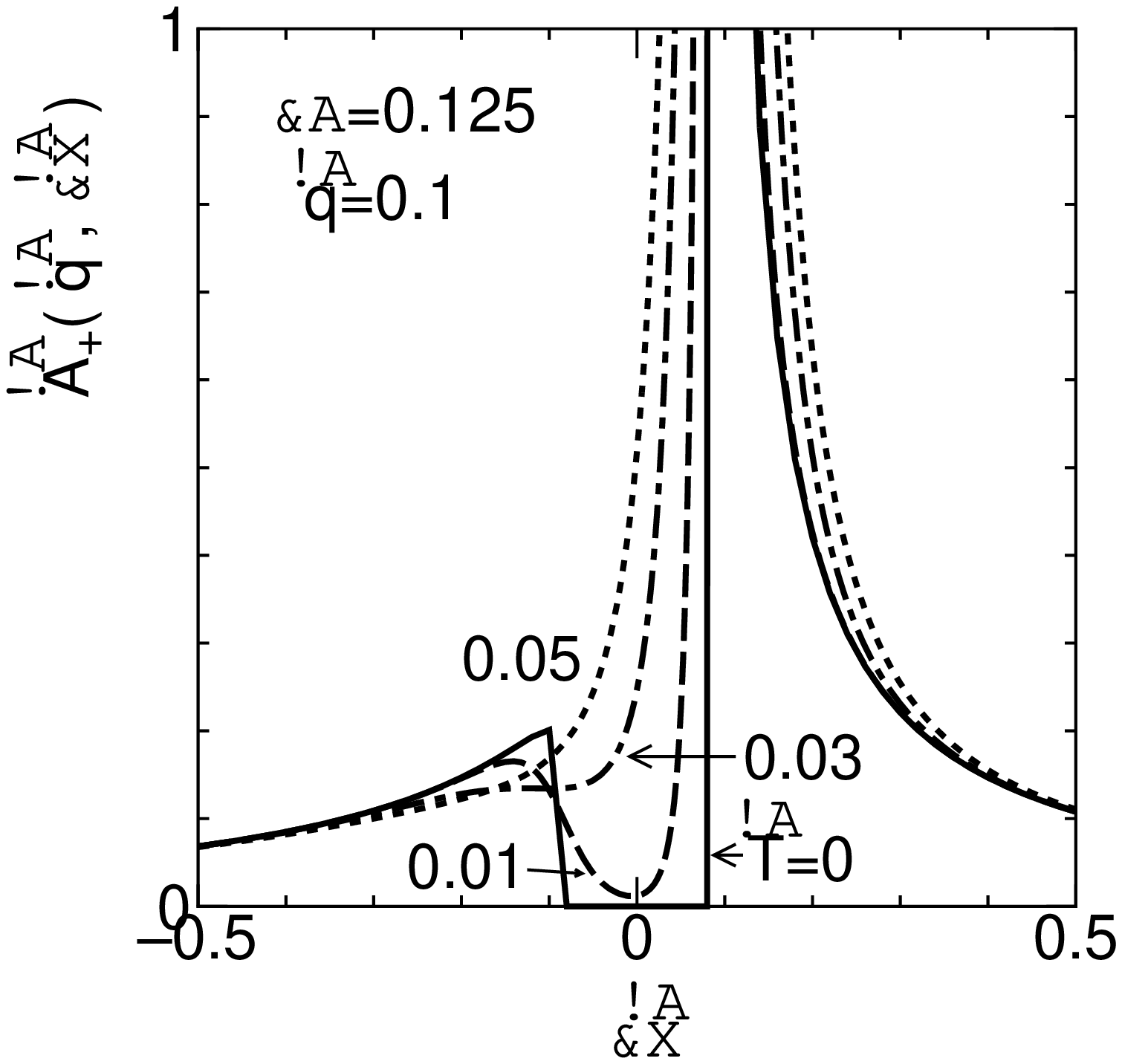}}
\vspace{-1.5cm}
%
\caption{
 The $\no$-dependence of 
 spectral function  $\nA$ 
  of the spinless model   
   where  $\nq=0.1$,  $\alpha=0.125$ and $\nT$ is chosen as   
  $\nT$=0, 0.01, 0.02, 0.05  and 0.1.  
}
\label{fig:3}
\end{wrapfigure}
%
 Next $\nA$ in case of  $\nq$ = 0.1 is examined. 
 The $\no$-dependence of $\nA$ in case of $\alpha = 0.125$ is shown in Fig. 3. 
 The quantity $\nA$ at $\nT=0$ shows not only the main peak 
 around $\no = \nv \nq$ but also that 
  near   $\no = - \nv \nq$ 
 where 
  $\nA \propto ( \no - \nv \nq)^{-1 + \alpha/2}$ 
for $\no \gsim \nv \nq$, 
    $\nA \propto | \no + \nv \nq|^{\alpha/2}$ 
for $ \no \lsim - \nv \nq$ and 
      $\nA = 0$ for $|\no| < \nv \nq$.\cite{Meden,Voit} 
 The peak around   $\no = - \nv \nq$ 
 comes from the particle-hole excitations between 
  two kinds of electrons with  $r=\pm$. 
 \cite{Voit} 
 The finite magnitude of $\nA$ appears 
in the interval region of 
 $|\no|<\nv \nq$ 
at finite temperatures. 
  By the increase of $\nT$, these two peaks are suppressed 
and merges into a single peak.

%
%
\subsection{spinful case}  
 We examine $\nA$ in the presence of   the interactions 
  $g_{2\parallel}$,   $g_{2\perp}$ and $g_{4 \perp}$ 
which result in the separation  of 
the charge degree of freedom from the spin degree of freedom. 
 When $\nq=0$, the $\no$-dependence of $\nA$ 
 is similar to Fig. 1 since the excitation spectra of 
 both charge fluctuation and spin fluctuation 
 become equal to zero at $\nq=0$. 
%
\begin{figure}[htb]
 \parbox{\halftext}{
\vspace{0.5cm}
\epsfxsize = 8.5cm
\vspace{-1.0cm}
\hspace{0.3cm}
\centerline{\epsfbox{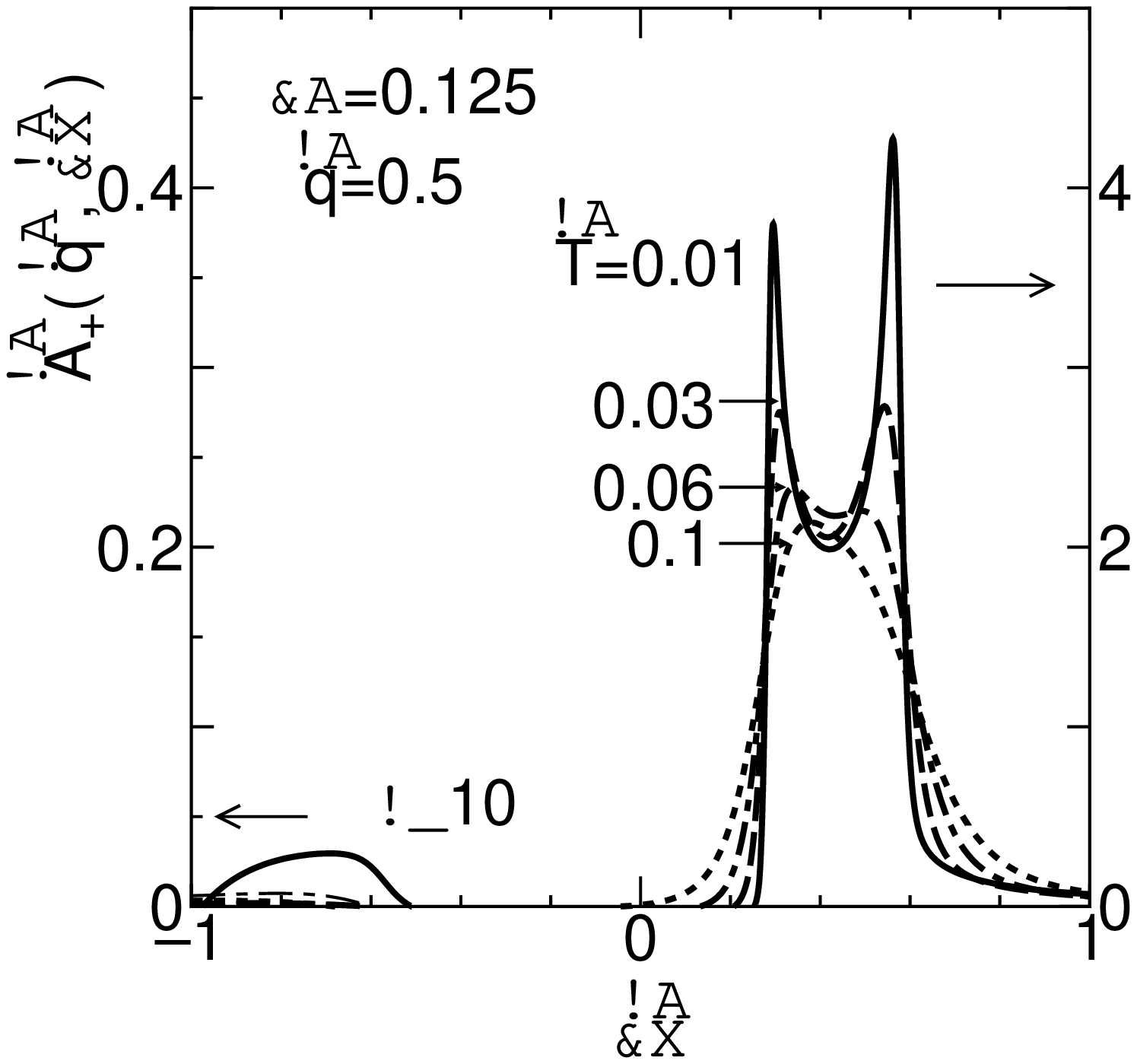}}
\vspace{-2.0cm}
\caption{
 The $\no$-dependence of 
 spectral function  $\nA$ 
  of the spinful  model   
   where  $\nq=0.5$,  $\alpha=0.125$ and $\nT$ is chosen as   
  $\nT$= 0.01(solid curve), 0.03(dashed curve), 
   0.06(dash-dotted curve) and 0.1(dotted curve).  
}
\label{fig:4}
}
\hspace{7mm}
\parbox{\halftext}{
\vspace{-1.3cm}
\epsfxsize = 8.5cm
\centerline{\epsfbox{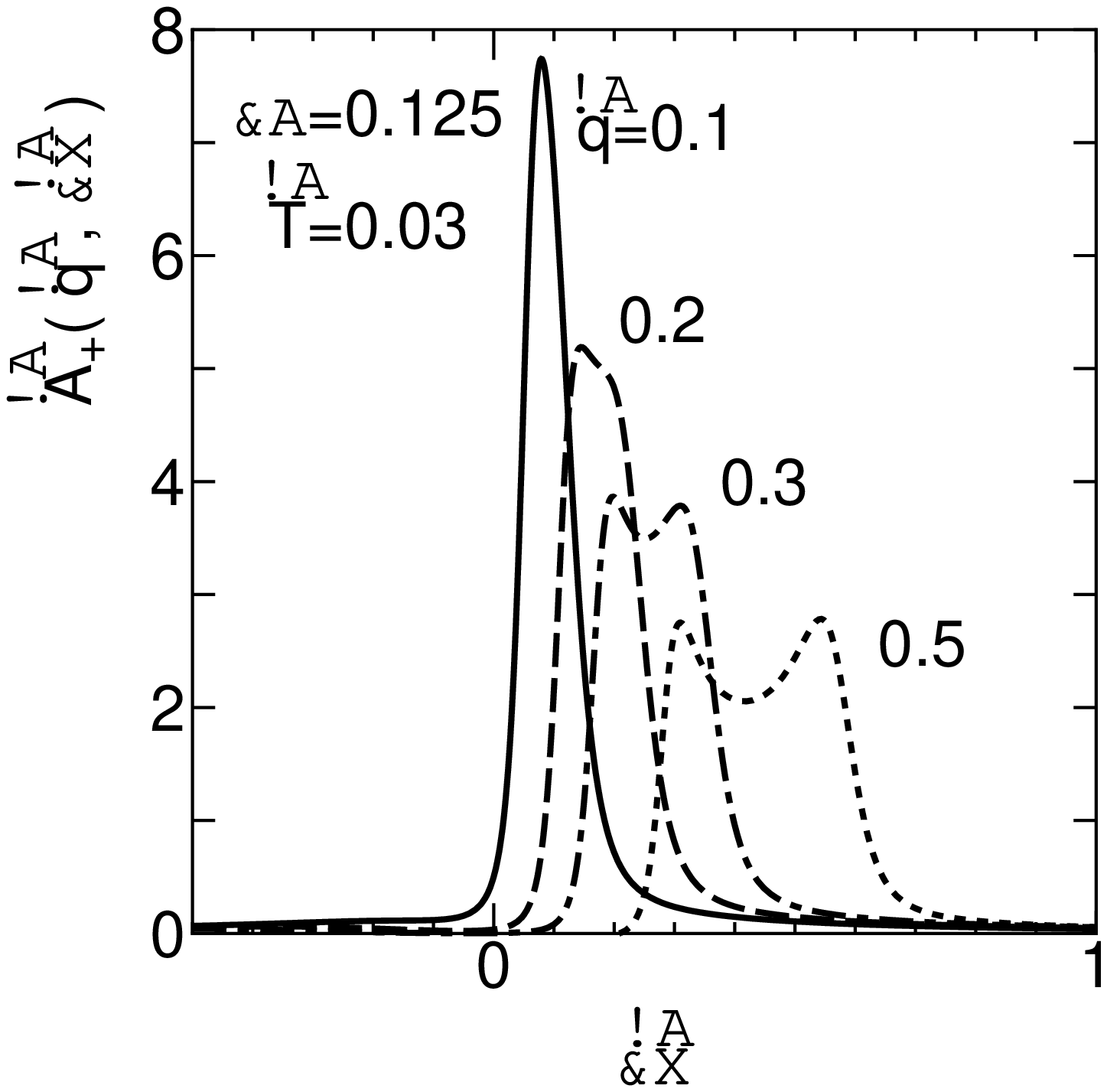}}
\vspace{-1.5cm}
\caption{
 The $\no$-dependence of 
 spectral function  $\nA$ 
  of the spinlful  model 
   where  $\alpha=0.125$, $\nT=0.03$   and $\nq$ is chosen as   
  $\nq$=0.1, 0.2, 0.3 and  0.5.  
}
\label{fig:5}
}
\end{figure}
%
 In Fig. 4,  $\nA$ in case of $\nq=0.5$ and $\alpha$=0.125   
 is shown with some choices of $\nT$ 
 where  $\nA$ with $\no < 0$ is multiplied by 10.   
  In case of  $\nT$ = 0, 
 there are several kinds of edges  in $\nA$ 
 which originate in  excitation spectra of 
   the spin and charge fluctuations. 
 Their asymptotic forms in the case of $\gas=0$  are given by
 \cite{Meden,Voit}
\begin{eqnarray} 
                                              \label{eqn:diverge}
 A_+(q,\omega)|_{\omega \simeq \vc q}
    &\propto& 
          |\omega-\vc q|^{\gac+2\gas-1/2}
 \virg \nonumber \\
 A_+(q,\omega)|_{\omega \simeq \vs q}
  &\propto&
  \step(\omega-\vs q)(\omega - \vs q)^{2\gac+\gas-1/2}
 \virg \nonumber \\
 A_+(q,\omega)|_{\omega \simeq - \vc q}
     &\propto& 
         \step(-\omega-\vc q) |\omega + \vc q|^{\gac+2\gas}
 \virg
\end{eqnarray}
 and $\nA = 0$ for $ - \vc \nq < \no < \vs \nq$ . 

 The $\nT$-dependence of $\nA$ is examined by defining  
  $\nAr$  ($\nAs$) as $\nA$ corresponding  to the peak  located 
 near $\omega = v_{\rho}q$ ($\omega = v_{\sigma}q$) 
 where $ v_{\sigma}q <  v_{\rho}q $. 
 The result  that  $\nAs <  \nAr$ in case of $\nT$ = 0.01 
can be understood from  the fact 
that, at $\nT$ = 0, 
the  exponent for the divergence of the charge excitation 
is larger than that  of the spin excitation 
as is seen from eqs. (\ref{eqn:diverge}). 
 By the increase of $\nT$, 
one finds that 
  $\nAs \simeq \nAr$ at $\nT$=0.03 
and that 
   $\nAs > \nAr$ for $\nT$=0.06. 
 Such a   crossover 
  from the dominant $\nAr$  into the dominant $\nAs$ 
  is  characteristic of the finite temperature. 
 The spin excitation,  which has the energy 
   lower than  that of the charge excitation,  
  has  the large  effect on   $\nA$ and 
   give rise to the dominant $\nAr$ 
   as is seen from  
    the interference term, 
     $\Omega_{\rho} s_{\sigma} + \Omega_{\sigma} s_{\rho}$, 
     in eq. (\ref{eqn:spectral2}).  
 The quantity $\nA$ at $\nT = 0.1$ shows that 
the two peaks  at low temperatures 
becomes  a single peak with the broad width. 
 When  $\nT$ increases, 
  $\nA$ with $\omega < 0$ 
decreases indicating the fact that 
the correlation by the interaction decreases 
by the thermal fluctuation. 
 In Fig. 5, we show $\nA$ with the fixed $\nT$ 
by choosing several $\nq$. 
 The interval length of $\no$  between  two peaks, 
which exists for the large $\nq$, 
decreases by the decrease of $\nq$ 
and vanishes for the small $\nq$, e.g., 
 $\nA$ for   $\no$ = 0.2 and 0.1. 
\newpage
%
\begin{wrapfigure}{r}{6.6cm}
\vspace{-0.5cm}
\epsfxsize= 8.5 cm
\vspace{-0.5cm}
\centerline{\epsfbox{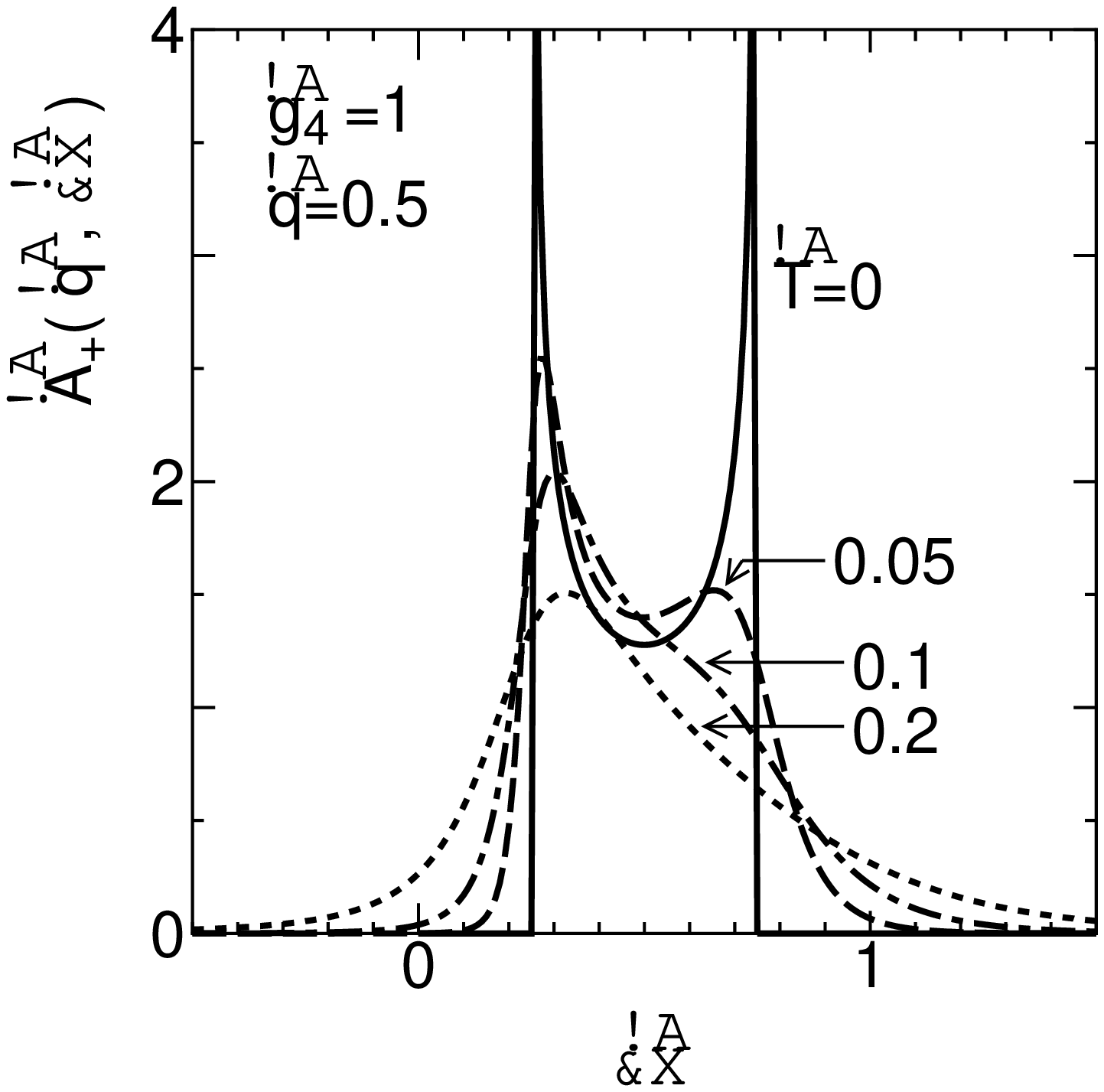}}
\vspace{-1.5cm}

\caption{
 The $\no$-dependence of 
 spectral function  $\nA$ 
  with the interaction,  $g_{4 \perp}$,    
   where  $\nq=0.5$,  $\widetilde{g}_{4 }= g_{4 \perp}/(\pi \vf)$      and $\nT$ is chosen as   
  $\nT$=0, 0.05, 0.1 and 0.2.  
}
\label{fig:6}
\end{wrapfigure}
%
For the comparison, 
$\nA$ with the interaction of only $g_{4 \perp}$ 
is shown in Fig. 6. 
 In case of $T=0$, $\nA$ exists only 
  in the interval region of  
  $v_{\sigma} q < \omega < v_{\rho} q$ 
   and show   two singularities  at the boundaries 
  corresponding to the spin and charge excitations 
  respectively.
  The exponents for the divergence at the charge excitation is 
the same as  the spin excitation 
where 
  $\nA \propto ( \no - \nvs \nq)^{-1/2}$ 
for $\no \gsim \nvs \nq$, 
  $\nA \propto (- \no + \nvc \nq)^{-1/2}$ 
for $ \no \lsim \nvc \nq$ and 
  $\nA = 0$ for $\no < \nvs \nq$ and $\no > \nvc \nq$ .
\cite{Voit} 
 The effect of the thermal fluctuation, 
  which leads to 
 the crossover from two peaks to the single peak, 
  is similar to  Fig. 5. 
 When  $\omega=\vs q$ or $\omega=\vs q$, 
 eq. (\ref{eqn:spectral}) is rewritten as 
\begin{eqnarray}
                                            \label{eqn:tem1}
 A_{+} (q, \vs q)
  &=& \frac{C_1}{T}
      \int_{0}^{\infty} \d y
      \frac{
       \cos \left( \frac{\vs q}{T} y \right)
      }
      {
       \left( \sinh y \right)^{\frac12}
      }
 \virg\\
                                            \label{eqn:tem2}
 A_{+} (q, \vc q)
  &=& \frac{C_1}{T}
      \int_{0}^{\infty} \d y
      \frac{
       \cos \left( \frac{\vc q}{T} y \right)
      }
      {
       \left( \sinh y \right)^\frac12
      }
 \virg
\end{eqnarray}
 where 
 $ C_1
  =  \sqrt{\vc \vs}
       \pi^{-2} |\vc-\vs|^{-1} 
      \int_{0}^{\infty} \d y
       \left( \sinh y \right)^{-\frac12}
 \point 
$
  Since 
  eqs. (\ref{eqn:tem1}) $\propto T^{-1/2} \vs^{-1}$  and 
  eqs. (\ref{eqn:tem2}) $\propto T^{-1/2} \vc^{-1}$   
  at low temperatures, 
  it turns out that 
  $\nAr$  decreases rapidly compared with $\nAs$.

%
\section{Discussion}
 We examined the spectral function, $\nA$, of 
 the Tomonaga-Luttinger model at finite temperatures 
for both the spinless case and the spinful case 
 by choosing   $\alpha=0.125$   which corresponds to  
  the  large  limit of the  repulsive interaction of 
  the Hubbard model.  
 In case of $\nq=0$, 
  $\nA$ shows the peak around $\no=0$.    
  By the increase of temperature, 
 the  height decreases  and the  width increases 
  due to the thermal fluctuation. 
  At low temperatures, 
  $\nA$   with  $\nq=0$ and $\no=0$, 
  shows the power law expressed as 
  $ \nA \propto \nT^{\alpha -1}$. 
 In case of $\nq >0 $, 
   there are two peaks  located in the region of   $\omega > 0$ 
    and that of $\omega < 0$ 
 for both the spinless and spinful cases.  
 With increasing temperature, 
 these two peaks  moves to a single peak around   $\omega = v q$.    In the spinful case, 
  the peak with $\nq >0$ is separated into two peaks 
   where  the peak with the large  $\no$,  $\nAr$,  corresponds 
    to the charge  fluctuation  
   and  the peak with the small  $\no$,  $\nAs$,  corresponds 
    to the spin  fluctuation.  
 In the limit of low temperatures, one obtains 
  $\nAs < \nAr$, while one finds 
 $\nAs > \nAs$      by the increase of temperature. 
    The latter result, which is explained   by   
      eq. (\ref{eqn:spectral2}), can be also understood  
  in terms of  the general formula
  \cite{Abrikosov} 
 \begin{equation}
                           \label{eqn:Amu}
 A (k , \omega)
  = \frac{1}{Z}
     \sum_{m,n}
     \left[
      |(C_k^+)_{n,m}|^2
       \e^{- \frac1{T}(E_{m}-\mu N_{m})}
        (
         1 + \e^{-\frac{\omega}{T}}
        )
      \delta ( E_n - E_m - \mu - \omega)
     \right], 
\end{equation}
  where  $Z=\sum_{n}\e^{(E_n-\mu N_n)/T}$ and 
   $N_n$ is the electron number.  
 Quantities   $E_n$ and $\mu$ are 
  the energy of the $n$-th eigenstate  
   and  the chemical potential 
   respectively  
 and  $(C_k^+)_{n,m}$ denotes the matrix element 
 between the $n$-th eigenstate and $m$-th eigenstate. 
   Since  $ \vc q > \vs q$ in the present case, 
   the factor,  $(1 + \e^{-\omega/T})$, with 
   $\omega = \vs q$   
     is larger than the factor with
   $\omega = \vc q$.  
   Therefore the peak for the spin fluctuation  becomes 
    larger than that for the charge fluctuation. 
 At higher temperatures, these two peaks  also merge 
  into a single peak. 

 Finally, we comment on the experiment 
 on  K$_{0.3}$MoO$_3$ where 
  the angle-resolved photoemission spectroscopy 
 reveals the two peaks corresponding to 
  the charge and spin separations.
  \cite{Gweon}
 As for  two peaks 
   indicating the properties of the Tomonaga-Luttinger liquid, 
 the peak with the lower energy is larger than 
 the peak with the higher energy. 
 In the Tomonaga-Luttinger model, 
the peak with lower energy is rather suppressed 
 when the interaction for the charge density is strong enough. 
 Then  Voit\cite{Voit_LE}
claimed that the peak with the higher energy is rather suppressed 
by the backward scattering.\cite{Luther-Emery}
 Here we comment  another possibility 
that the suppression of the peak with higher energy 
is attributable to the effect of thermal fluctuation 
as is found  in Figs. 4 and 6. 
\section*{Acknowledgements}

 The authors are thankful to  H. Yoshioka, T. Matsuura  and Y. Kuroda 
  for useful discussion.  
%
%

%
%
%

\begin{thebibliography}{99}
\bibitem{Tomonaga}
   S. Tomonaga,
   Prog. Theor. Phys. {\bf 5} (1950), 544.
\bibitem{Luttinger}
   J. M. Luttinger,
   J. Math. Phys. {\bf 4} (1963), 1154.
\bibitem{Luther}
   A. Luther and I. Peschel,
   Phys. Rev.\ {\bf B9} (1974), 2911.
\bibitem{Solyom}
   J. S\'olyom,
   Adv. Phys. {\bf 28} (1979), 201.
\bibitem{Mattis}
   D. C. Mattis and E. H. Lieb,
   J. Math. Phys. {\bf 6} (1965), 304.
\bibitem{Suzumura_P2}
   Y. Suzumura,
   Prog. Theor. Phys. {\bf 63} (1980), 51.
\bibitem{Nakamura}
   N. Nakamura and Y. Suzumura,
   Prog. Theor. Phys. {\bf 97} (1997), 163.
\bibitem{Voit}
   J. Voit,
   Phys. Rev. B {\bf 47} (1993) 6740; 
   J. of Phys. : Condens. Matter {\bf 5} (1993), 8305.
\bibitem{Haldane}
   F.D. Haldane, 
   J. Phys. C: Solid State Phys. {\bf 14} (1981), 2585.
\bibitem{Meden}
   V. Meden and K. Sch$\ddot{\rm{o}}$nhammer,
   Phys. Rev. B {\bf 46} (1992) 15753;
     K. Sch$\ddot{\rm{o}}$nhammer and V. Meden,
   Phys. Rev. B {\bf 47} (1993) 16205;
\bibitem{Penc2}
   K. Penc, K. Hallberg, F. Mila and H. Shiba,
   Phys. Rev. Lett. {\bf 77} (1996), 1390.
\bibitem{Suzumura_P1}
   Y. Suzumura:
   Prog. Theor. Phys. {\bf 61} (1979) 1.
\bibitem{Dzyaloshinskii}
   I. E. Dzyaloshinskii and I. Larkin,
   Sov. Phys.-JETP {\bf 38} (1974), 202.
\bibitem{Ogata}
   M. Ogata and P.W. Anderson, 
   Phys. Rev. Lett.\ B {\bf 70} (1993), 3087. 
\bibitem{Schulz}
   H. J. Schulz,
   Phys. Rev. Lett. {\bf 64} (1990), 2831.
\bibitem{Abrikosov}
   A.A. Abrikosov, L.P. Gorkov and I.E. Dzyaloshinskii,
   {\it
 Method of Quantum Field Theory in Statistical Physics   
   }
   (Printice-Hall, 1963).
\bibitem{Gweon}
   G. H. Gweon et al,
   J. of Phys. : Condens. Matter {\bf 8} (1996), 9923.
\bibitem{Voit_LE}
   J. Voit:
   J. of Phys. : Condens. Matter {\bf 8} (1996), L779.
\bibitem{Luther-Emery}
   A. Luther and V.J. Emery,
   Phys. Rev. Lett. {\bf 33} (1974), 589.
%
%
\end{thebibliography}
\end{document}